\documentclass{article}

\def\abstract#1{\vskip 7mm 
        \begin{center}{\large Abstract}\par \smallskip
                \begin{minipage}[c]{12cm}
                        \small #1
                \end{minipage}
        \end{center}
}
\def\title#1{\begin{center}{\Large\bf #1}\end{center}}
\def\author#1{\vskip 5mm \begin{center}{#1}\end{center}}
\def\address#1{\begin{center}{\it #1}\end{center}}
\makeatletter
\@ifundefined{lesssim}{}{}
\@ifundefined{gtrsim}{}{}
\def\vereq#1#2{\lower3pt\vbox{\baselineskip1.5pt \lineskip1.5pt
\ialign{$\m@th#1\hfill##\hfil$\crcr#2\crcr\sim\crcr}}}
\makeatother

\begin{document}

\title{
  Mass and angular momenta of Kerr-anti-de Sitter spacetimes
  \smallskip \\}
\author{
  Nathalie Deruelle\footnote{deruelle@ihes.fr}}
\address{
 Institut des Hautes Etudes Scientifiques,
35 Route de Chartres, 91440, Bures-sur-Yvette,
France}

\abstract{
 We briefly review how to compute the mass and angular momenta of rotating,
asymptotically anti-de Sitter spacetimes in
 Einstein-Gauss-Bonnet theory of gravity using superpotentials derived
from standard N\oe ther identities. The calculations depend on the asymptotic form of
the metrics only and hence take no account of the source of the curvature. When the
source of curvature is a black hole, the results can be checked using the first law of
black hole thermodynamics.}

\section{Introduction}

Consider  Kerr-AdS$_4$ spacetime (${\cal M}$) in Boyer-Lindquist coordinates 
[1] 
$$ds^2=-{\Delta\over\rho^2}\left(dt-{a\,\sin^2\theta\over\Xi}\,
d\phi\right)^2+{\rho^2\over\Delta} dr^2+{\rho^2\over
\Delta_\theta}d\theta^2+{{
\Delta_\theta}\sin^2\theta\over\rho^2}\left(a\,dt
-{r^2+a^2\over\Xi}d\phi\right)^2$$
where
$\Delta\equiv(r^2+a^2)(1+r^2/l^2)-2m\,r\,$, $\,\rho^2\equiv r^2+a^2\cos^2\theta\,$,
$\,\Delta_\theta\equiv1-(a^2/ l^2)\cos^2\theta$ and $\Xi\equiv1-a^2/l^2$. It solves
the Einstein equations with cosmological constant,
$R^\mu_\nu=-(3/l^2)\delta^\mu_\nu$, and depends on two integration constants, the
``mass" and  ``rotation" parameters $m$ and
$a$. 
${\cal M}$
 is asymptotically anti-de Sitter but, as noticed in [2], the Boyer-Lindquist
frame is rotating with angular velocity $\Omega_{BL}=a/l^2$ with respect to the
standard, static,  AdS$_4$ frame.

${\cal M}$ describes a black hole. Its horizon $r_+$, which can be defined as the
surface where signals from infalling test particles are measured at infinity with
infinite redshift by static observers, is located at
$$\Delta(r_+)=0\quad\Longleftrightarrow\quad
m={(r_+^2+a^2)(1+r_+^2/l^2)\over2r_+}\,.\quad$$
Its angular velocity can be defined as the angular velocity of photons
orbiting the horizon, as measured at infinity with respect to the static  AdS$_4$
frame, and is
$$\Omega={a(1+r_+^2/l^2)\over r_+^2+a^2}\,.$$

A Bekenstein entropy can be associated to this black hole, as well as a
Hawking black body temperature, respectively related to the area $A$ of the horizon
and its surface gravity $\kappa$ by
$$S={1\over4} A={\pi\,(r^2_++a^2)\over\Xi}\quad,\quad T={\kappa\over
2\pi}= {4\pi\,(r^2_++a^2)\over r_+(1+a^2/l^2+3r_+^2/l^2-a^2/r_+^2)}\,.$$

Since ${\cal M}$ is stationary and axially symmetric, it possesses two Killing vectors.
Hence there exists global conserved charges that can be interpreted as the mass $M$
and angular momentum $J$ of the black hole. There are various ways to construct them.
As reviewed in [2] there is general agreement in the literature on the
definition of
$J$ but not on the definition of $M$. The ``right" results are
$$M={m\over\Xi^2}\quad,\quad J={ma\over\Xi^2}$$
where the mass is associated with the Killing vector describing time translations in a
locally free falling {\it non rotating} inertial frame in AdS$_4$. In the rotating Boyer-Linquist frame its
components are
$(1,0,0,-a/l^2)$.

The reason why one can speak of the ``right" results for $M$ and $J$ is that, as
emphasized in [2], the first law of thermodynamics must hold, that is, one
must have
$$ T\, dS=dM-\Omega\, dJ\,.$$
Now, $T,S,\Omega,M,J$ are known functions of the two parameters $m$ and $a$ (or,
computationally more simply, of $r_+$ and $a$). Therefore the two equalities
$$T{\partial S\over\partial r_+}={\partial M\over\partial r_+}-\Omega{\partial
J\over\partial r_+}\quad,\quad T{\partial S\over\partial
a}={\partial M\over\partial a}-\Omega{\partial J\over\partial a}$$
must hold and, if they do, provide a consistency check of the interpretation of
the Kerr-AdS$_4$ metric as representing a radiating black hole spacetime, and in
particular a check of the pertinence of the definition of its mass.

The object of this contribution is, first, to review in section 2 a method based on
standard N\oe ther identities which
defines conserved charges associated to a spacetime ${\cal M}$ with respect to its
asymptotic background $\overline{{\cal M}}$ in Einstein-Gauss-Bonnet (EGB) theory
[3], irrespective of the source of curvature, that is whether it is a ``star" or a black
hole.

In Einstein's theory, when applied to
 the recently discovered D-Dimensional Kerr-anti-de Sitter Einstein
black holes [4], that method yields the right mass, that is, for which the first
law holds, as shown in [5] and
reviewed in Section 4. 

In Einstein-Gauss-Bonnet's theory the rotating Kerr-like solution  is not known.
However, on one hand, only the asymptotic form of the metric is required to define
the conserved charges and, on the other hand, the Einstein and
Einstein-Gauss-Bonnet vacuum equations are the same when linearized on a
maximally symmetric background (see section 3) : hence the
D-dimensional Kerr-anti-de Sitter solutions to the Einstein field equations found in
[4] are also asymptotic solutions to the Einstein-{\it Gauss-Bonnet}
equations. Those two remarks were used in [6]
 to give an explicit expression for the mass and
angular momenta of Kerr-AdS spacetimes in EGB theory, which is reviewed in section
4. When the full geometry of the Kerr-like rotating EGB black holes is eventually
obtained, a test of the proposed definitions will be to see if  the first law of EGB
black hole thermodynamics is satisfied.

\section{N\oe ther superpotential and charges in EGB theory}

There are various ways to associate conserved charges to a $D$-dimensional
spacetime
${\cal M}$. We shall review here the approach advocated in [7] and [8] and
applied to  Einstein-Gauss-Bonnet theory of gravity in [3].

Consider the lagrangian
$${\cal L}=L+D_\mu k^\mu\eqno(2.1)$$
 where $L$ is some function of curvature invariants, functional of
the metric $g_{\mu\nu}(x^\rho)$, its first and second derivatives with respect to
the coordinates $x^\rho$ and where 
$k^\mu$ is a vector to be appropriatly chosen (see below). 
 The variation of the scalar  density $\hat{\cal L}$ with respect to the
metric can be written  as
$$\delta\hat{\cal L}=-\hat\sigma^{\mu\nu}\delta g_{\mu\nu}+
\partial_\mu(\hat V^\mu+\delta \hat k^\mu)\,. \eqno(2.2)$$
 (A
hat denotes multiplication by $\sqrt{-g}$ where $g$ is the determinant of the
metric.) 
The tensor $\sigma^{\mu\nu}$ is the variational derivative of
$\hat{\cal L}$ and the field equations  are $\sigma^{\mu\nu}=0$, whatever the vector
$k^\mu$. As for the vector $V^\mu$ it can be written under the form
$$V^\mu=\alpha^{\mu\nu\rho}\,\delta g_{\nu\rho}+
\beta^{\mu\nu\rho}_{\hskip 0.5cm \sigma}\
\delta\Gamma^\sigma_{\nu\rho}\eqno(2.3)$$  
where $\Gamma^\sigma_{\nu\rho}$ are the
Christoffel symbols, where 
$\delta\Gamma^\sigma_{\nu\rho}={1\over 2}g^{\sigma\lambda}(D_\nu\delta
g_{\rho\lambda}+D_\rho\delta g_{\nu\lambda}-D_\lambda\delta g_{\nu\rho})$ and
where the tensors $\alpha^{\mu\nu\rho}$ and $\beta^{\mu\nu\rho}_{\hskip 0.5cm
\sigma}$ depend on the precise dependence of $L$ on the metric and its derivatives.  

If the variation $\delta g_{\mu\nu}$ is due to a mere
 change of coordinates $x^\mu\to\tilde x^\mu=x^\mu-\xi^\mu$ with $\xi^\mu$ an
infinitesimal vector (so that $\delta\hat{\cal L}= \partial_\mu(\hat{\cal
L}\xi^\mu)$ and $\delta
g_{\mu\nu}=2D_{(\mu}\xi_{\nu)}$, parentheses denoting
symmetrisation), then the variation (2) of the Lagrangian density  can be cast in the
form~:
$$\partial_\mu\hat j^\mu=D_\mu\hat j^\mu=
0\qquad\hbox{with}\qquad \hat j^\mu\equiv(2\hat\sigma^{\mu\nu}+ g^{\mu\nu}\hat
L)\xi_\nu-\hat V^\mu+2D_\nu(\xi^{[\mu}\hat k^{\nu]})\eqno(2.4)$$
thanks to the (generalized) Bianchi identities. (Brackets stand for
anti-symmetrization : $f_{[\mu\nu]}\equiv {1\over2}(f_{\mu\nu}-f_{\nu\mu})$.)

The conservation of $j^\mu$ implies that there exists an antisymmetric
 tensor $j^{[\mu\nu]}$ such that 
$$\hat j^\mu\equiv D_\nu\hat j^{[\mu\nu]}=\partial_\nu\hat
j^{[\mu\nu]}\,.\eqno(2.5)$$ 
If we look  for an expression of the form
$$j^{[\mu\nu]}={\cal A}^{[\mu\nu]\rho}\xi_\rho+{\cal
B}^{[\mu\nu]\rho\sigma}D_\rho\xi_\sigma+2\xi^{[\mu} k^{\nu]}\eqno(2.6)$$  
then ${\cal A}^{[\mu\nu]\rho}$ and ${\cal B}^{[\mu\nu]\rho\sigma}$ are obtained in
terms of $\alpha^{\mu\nu\rho}$ and $\beta^{\mu\nu\rho}_{\hskip 0.5cm
\sigma}$  from (4) and (5) by identification.

We obtain similarly $\overline{\hat j^\mu}$ and $\overline{\hat j^{[\mu\nu]}}$ for the
background $\overline{{\cal M}}$ (an 
overline means that the quantity is evaluated on the background  spacetime with
metric  $\overline{g}_{\mu\nu}(x^\rho)$). In
practice the only demand on $\overline{{\cal M}}$ is that it asymptotes ${\cal M}$. The
KBL superpotential density [7] [8] is then  defined as 
$$\hat J^{[\mu\nu]}=-{1\over16\pi} \left(\hat j^{[\mu\nu]}-\overline{\hat
j^{[\mu\nu]}}\right)\,.\eqno(2.7)$$

We now integrate (7) over a region ${\cal M}$ of spacetime
bounded by
$\partial{\cal M}=\partial\overline{{\cal M}}$, a timelike hypercylinder limited by two
spacelike hypersurfaces 
$x^0=t_0$ and $x^0=t_1$. We apply Stokes' theorem 
and obtain that
 the ``charge"
$$Q=\int_S\!d^{D-2}x\ n_i\hat J^{[0i]}\eqno(2.8)$$
 is independent of time if $\int_S\!d^{D-2}x\ n_i(\hat j^i-\overline{\hat j^i})=0$
where 
$S$ is the
$(D-2)$-dimensional sphere at infinity, with $n^i$ its unit normal vector pointing
towards infinity.

 In practice
 we shall restrict our attention  to stationary  spacetimes and
choose $\xi^\mu$ to be a Killing vector, so that $Q$ will indeed be time-independent. When that Killing
vector is associated with time translations with respect to a non rotating frame at
infinity the associated conserved charge is the mass $M$ of ${\cal M}$. 
If  the Killing vector is associated with rotations then the associated charges are the
angular momenta $J_i$. 

As for 
the vector $k^\mu$ it is chosen according to circumstances (that is, according to  $L$)
in order to cancel as many
 normal derivatives of the  metric variations as possible in the divergence in (2). If
they all cancel, the field equations are obtained by keeping only the metric and its
tangential derivatives fixed at the boundary, leaving their normal derivatives free,
and the variational problem falls within the scope of ordinary Lagrangian  field
theory.

\bigskip

Let us now specialize to Einstein-Gauss-Bonnet's theory of gravity. The
 Lagrangian is (see e.g. [9] for a review of its properties)
 $$L\equiv -2\Lambda+R+\alpha\,
R^{\mu\nu\rho\sigma}P_{\mu\nu\rho\sigma}\quad \hbox{with}\quad
P_{\mu\nu\rho\sigma}\equiv R_{\mu\nu\rho\sigma}-2(R_{\mu[\rho}g_{\sigma]\nu}
-R_{\nu[\rho}g_{\sigma]\mu})+R \, g_{\mu[\rho}g_{\sigma]\nu}\eqno(2.9)$$
(so that $R^{\mu\nu\rho\sigma}P_{\mu\nu\rho\sigma}=
R^{\mu\nu\rho\sigma}R_{\mu\nu\rho\sigma}-4R^{\mu\nu}R_{\mu\nu}
+R^2)\,$.
$\Lambda$ is a (negative) cosmological constant, $\alpha$ a coupling constant
which is zero in Einstein's theory, and $R_{\mu\nu\rho\sigma}$, $R_{\mu\nu}$
and
$R$ are the Riemann tensor, Ricci tensor and curvature scalar of the metric
$g_{\mu\nu}$. 

The equations of motion derived from that
Lagrangian are
$$\sigma^{\mu\nu}\equiv\sigma^{\mu\nu}_{E}+\alpha\sigma^{\mu\nu}_{GB}=0
\eqno(2.10)$$ with
$$\sigma^{\mu\nu}_{E}\equiv\Lambda
g^{\mu\nu}+R^{\mu\nu}-{1\over2}R\,g^{\mu\nu}\eqno(2.11)$$ and 
$$\sigma^{\mu\nu}_{GB}\equiv 2R^{\mu\lambda\rho\sigma}P^\nu_{\ \ \lambda
\rho\sigma}-{1\over2}g^{\mu\nu}
R^{\alpha\beta\rho\sigma}P_{\alpha\beta\rho\sigma}\,.\eqno(2.12)$$

The explicit form of the vector $V^\mu$ appearing in (2) (3) is 
$$V^\mu\equiv V^\mu_{E}+\alpha V^\mu_{GB}\eqno(2.13)$$
with
$$V_{E}^\mu=-\left(g^{\mu\nu}\delta\Gamma^\rho_{\nu\rho}-g^{\nu\rho}
\delta\Gamma^\mu_{\nu\rho}\right)\eqno(2.14)$$
and
$$V_{GB}^\mu=4D^{(\mu}\sigma_{E}^{\nu)\rho}\delta
g_{\nu\rho}-4\left(P^{\mu\alpha\beta}_{\ \ \ \ \
\gamma}-Q^{\mu\alpha\beta}_{\ \ \ \ \
\gamma}\right)\delta\Gamma^\gamma_{\alpha\beta}\eqno(2.15)$$
where the tensor $P_{\mu\nu\rho\sigma}$ is defined in (9) and where
$Q_{\mu\nu\rho\sigma}\equiv2(R_{\rho[\mu}g_{\nu]\sigma}+
R_{\sigma[\mu}g_{\nu]\rho})$. 
 The tensors ${\cal A}^{[\mu\nu]\rho}$ and ${\cal
B}^{[\mu\nu]\rho\sigma}$ in (6) are then easily calculated and yield the following
expression for the EGB superpotential
$$\hat J^{[\mu\nu]}\equiv\hat J^{[\mu\nu]}_E+\alpha\, \hat
J^{[\mu\nu]}_{GB}\eqno(2.16)$$
where the
Einstein contribution  is given by, see [7-8] 
$$-8\pi \hat
J^{[\mu\nu]}_E\equiv
D^{[\mu}\hat\xi^{\nu]}-\overline{D^{[\mu}\hat\xi^{\nu]}}+\hat\xi^{[\mu}k_E^{\nu]}
\,.\eqno(2.17)$$
As for the Gauss-Bonnet contribution $\hat J^{[\mu\nu]}_{GB}$ it is, when $\xi^\mu$
is a Killing vector [3]
$$-8\pi \hat
J^{[\mu\nu]}_{GB}\equiv2
\left[P^{\mu\nu\alpha\beta}D_{[\alpha}\hat\xi_{\beta]}-
\overline{P^{\mu\nu\alpha\beta}D_{[\alpha}\hat\xi_{\beta]}}\right]
+\hat\xi^{[\mu}k_{GB}^{\nu]}\,.\eqno(2.18)$$

The last step is to choose the vectors $k^\mu_E$ and $k^\mu_{GB}$. The choice made in
[7] and [8], that is
$$k^\nu_E=
g^{\mu\nu}\Delta^\rho_{\nu\rho}-g^{\nu\rho}\Delta^\mu_{\nu\rho}
\quad\hbox{with}\quad\Delta^\mu_{\nu\rho}\equiv\Gamma^\mu_{\nu\rho}
-\overline{\Gamma^\mu_{\nu\rho}}\eqno(2.19)$$
has passed all tests with flying colours (see e.g. [10]). The following proposal was
made in [3] for the Gauss-Bonnet contribution
 $$k^\mu_{GB}=4\left(P^{\mu\alpha\beta}_{\ \ \ \ \
\gamma}-Q^{\mu\alpha\beta}_{\ \ \ \ \
\gamma}\right)\Delta^\gamma_{\alpha\beta}\eqno(2.20)$$
where it was shown that such a definition yielded the correct formula for the
mass of the static EGB-Schwarzschild-like spacetimes, that is, which is consistent
with the first law of EGB black hole thermodynamics. 

More generally, when the vector $\xi^\mu_t$ is the Killing vector
of time translations with respect to a non rotating frame at infinity
the associated conserved charge is the mass $M$. If the vector
$\xi^\mu_i$ is a Killing vector for rotations with respect to the $i$-axis
 then the associated charges is the angular momentum $J_i$ :
$$M=\int_Sd^{D-2}x\hat J^{[01]}_t\quad,\quad J_i=\int_Sd^{D-2}x\hat
J^{[01]}_i\,.\eqno(2.21)$$ 
 
Equipped with those formulae it is then a matter of straightforward calculation to
obtain the mass $M$ and angular momenta $J_i$ for a given metric which solves the
field equations (10-12). It is important to note  that only the asymptotic 
form of the metric is needed and that the mass and angular momenta thus obtained do
not depend on the source of curvature, which can be a ``star" or a black hole.

\section{The asymptotic Kerr-AdS solutions of  the EGB equations}

The Einstein-Gauss-Bonnet equations of motion were given in the previous section,
equations (2.10-2.12). In order to compute the mass and angular momenta of stationary
and axi-symmetric spacetimes which solve those field equations, only the aymptotic
form of the metric is required. 
We therefore linearize the EGB equations, that is, we set
 $$g_{\mu\nu}=\overline{g}_{\mu\nu}+h_{\mu\nu}\,.\eqno(3.1)$$

 We
choose the background $\overline{g}_{\mu\nu}$ to be the metric of a maximally
symmetric spacetime $\overline{\cal M}$ solving the EGB equations. Since
they are quadratic in the Riemann tensor there are two solutions, defined by
$$\overline{R}_{\mu\nu\rho\sigma}=-{1\over{\cal
L}^2}(\overline{g}_{\mu\rho}\overline{g}_{\nu\sigma}-
\overline{g}_{\mu\sigma}\overline{g}_{\nu\rho})
\quad\hbox{where}\quad {1\over{\cal
L}^2}\equiv{1\over2\tilde\alpha}\left(1\pm\sqrt{1-{4\tilde\alpha\over
l^2}}\right) \eqno(3.2)$$ 
with $\tilde\alpha\equiv(D-3)(D-4)\alpha$ and
$l^2\equiv-{(D-1)(D-2)\over2\Lambda}\ $, $D$ being the dimension of spacetime. We
shall choose the lower sign for ${\cal L}^2$ so that the limit when $\alpha=0$
solves the Einstein equations. (When $\Lambda$ and $\alpha$ are such that
$\sqrt{1-4\tilde\alpha/ l^2}=0$ the two roots coalesce into a single double
root.)

Using the fact that
$$R^\mu_{\ \nu\rho\sigma}=\overline{R^\mu_{\ 
\nu\rho\sigma}}+{1\over2}\overline{D}_\rho(\overline D_\nu
h^\mu_\sigma+\overline{D}_\sigma h^\mu_\nu-\overline{D}^\mu
h_{\sigma\nu})-{1\over2}\overline{D}_\sigma(\overline D_\nu
h^\mu_\rho+\overline{D}_\rho h^\mu_\nu-\overline{D}^\mu
h_{\rho\nu})+{\cal O}(h^2)\eqno(3.3)$$ 
($\overline{D}_\mu$ being the covariant derivative associated to
$\overline{g}_{\mu\nu}$)  it is an exercise to compute $R^{\mu\lambda\rho\sigma}P_{\nu\lambda
\rho\sigma}$ at linear order and see
that the EGB equations  read (a result first obtained in [11] for
$\Lambda=0$) 
$$ \sqrt{1-{4\tilde\alpha\over
l^2}}\left[ R^\mu_\nu-{1\over2}\delta^\mu_\nu\, R+\delta^\mu_\nu\,\Lambda
{l^2\over{\cal L}^2}\right]={\cal O}(h^2)\eqno(3.4)$$ 
where
$$R^\mu_\nu=-{(D-1)\over{\cal
L}^2}(\delta^\mu_\nu-h^\mu_\nu)+
{1\over2}\left[\overline{D}_\alpha(\overline{D}_\nu
h^{\alpha\mu}+\overline{D}^\mu h_\nu^\alpha-\overline{D}^\alpha
h^\mu_\nu) -\overline{D}^\mu\overline{D}_\nu h^\rho_\rho\right]+{\cal
O}(h^2)\eqno(3.5)$$
 is the linearized Ricci tensor on $\overline{\cal M}$. 

Hence, when
linearized on a maximally symmetric background, the Einstein-Gauss-Bonnet equations
are the same as the linearized pure Einstein equations  $(\alpha=0)$ with effective
cosmological constant
$\Lambda_e\equiv\Lambda{l^2\over{\cal L}^2}$, and up to the overall factor
$\sqrt{1-4\tilde\alpha/l^2}$ that we shall assume not to be zero. 
\bigskip

Now, in [4] Gibbons, Lu, Page and Pope found D-dimensional Kerr-anti-de
 Sitter exact solutions of Einstein's equations. Since the
linearized Einstein-Gauss-Bonnet equations are the same as the linearized pure
Einstein equations,
those solutions  also solve the Einstein-Gauss-Bonnet equations at linear order.

We shall write them  in Kerr-Schild
coordinates. In odd dimensions $D=2n+1$ for instance, the line element reads
$$
ds^2=g_{\mu\nu}dx^\mu dx^\nu=d\overline{s}^2 +{2m\over U}\,(h_\mu
dx^\mu)^2\qquad {\rm with}\quad \mu=\{0, 1, \mu_i, \phi_i\}\quad (i=1,...,n)
\eqno{(3.6)}
$$  
where $d\overline{s}^2$ is the metric of the anti-de Sitter background and where the
coordinates $\mu_i$ are subject to the constraint
$\sum_{i=1}^{i=n}\mu_i^2=1$.  The mass parameter $m$ is an integration constant, 
$h_\mu$ is some null vector (denoted $k_\mu$ in [2] [4]) depending on
$n$ rotation parameters $a_i$ and  $U$ is a scalar function. Explicit forms of those
functions can be found in Section 2 of ref [4]. 

As
an example, the 5-dimensional AdS line element
in Kerr-Schild
ellipsoidal coordinates $x^\mu=(t,r,\theta,\phi,\psi)$ reads :
$$d\overline{s}^2=-{(1+r^2/{\cal
L}^2)\Delta_\theta\over\Xi_a\Xi_b}dt^2+ {r^2\rho^2\over(1+r^2/{\cal
L}^2)(r^2+a^2)(r^2+b^2)}dr^2+{\rho^2\over\Delta_\theta}d\theta^2+$$
$$+{r^2+a^2\over\Xi_a}\sin^2\theta\, d\phi^2+{r^2+b^2\over\Xi_b}\cos^2\theta\,
d\psi^2\eqno(3.7)$$
 where $\Delta_\theta\equiv
\Xi_a\cos^2\theta+\Xi_b\sin^2\theta$, where $\rho^2\equiv r^2+ {\cal
L}^2(1-\Delta_\theta)$, and where $\Xi_a$ and $\Xi_b$ are related to the rotation
parameters $a$ and $b$ by $\Xi_a\equiv 1-a^2/{\cal L}^2$, $\Xi_b\equiv
1-b^2/{\cal L}^2$. As for the function $U$ and the null vector $h_\mu$ they are
given by $U=\rho^2$ and 
$$h_\mu
dx^\mu={\Delta_\theta\over\Xi_a\Xi_b}\,dt+{r^2\rho^2\over(1+r^2/{\cal
L}^2)(r^2+a^2)(r^2+b^2)}\,dr+
{a\sin^2\theta\over\Xi_a}d\phi+{b\cos^2\theta\over\Xi_b}d\psi\,.\eqno(3.8)$$
 Note
that we shall only need their asymptotic behaviours at leading order. Note also that
the full Ricci tensor of the Kerr-Schild metric (6-8) is linear in
$h_{\mu\nu}={2m\over U} h_\mu h_\nu$ and hence exactly given by (5), see e.g.  [4].

The background admits a timelike Killing vector $\xi=\xi_t$ and $n$ plane rotation
Killing vectors $\xi=\xi_i$ associated with one parameter
displacements, say $\tau$.
$\xi_t$ is normalized in such a way that $\xi^\mu_t\delta\tau$   is an infinitesimal
time translation in a locally free falling {\it non rotating} inertial frame in the
background. Thus for the background metric in its conventional static form when
$D=2n+1$ for instance (equations (2.11) and (2.12) of Ref [4])
$$
 d\bar s^2= -\left(1+ {y^2\over l^2}\right)dt^2+{dy^2\over{1+{y^2/ l^2}}}+y^2 
d\sigma^2_{(D-2)}\qquad\hbox{where}\quad  d\sigma^2_{(D-2)}=
\sum_{i=1}^{i=n}(d\tilde\mu_i^2+\tilde\mu_i^2 d\phi_i^2),
\eqno{(3.9)}
$$ 
with $\Sigma_i\tilde\mu_i^2 =1$,
 the components of the timelike Killing vector $\xi_t$ corresponding to time
translations are  $ (1,0,0,0,...)$. In Kerr-Schild ellipsoidal coordinates
$\xi_t$ has   the same components. If the vector
$\xi^\mu_i$ is the Killing vector associated with rotations, for example with respect
to the
$\phi$ (resp. $\psi$) axis in five dimensions, that is $\xi_a^\mu=(0,0,0,1,0)$ (resp.
$\xi_b^\mu=(0,0,0,0,1)$), then the associated charges are the angular momenta.

\section{Kerr-AdS masses and angular momenta in EGB theory}

To obtain the mass and angular momenta of the 5D-Kerr-like solution of
 EGB theory we inserted the metric (3.6-3.8) of section 3 in the definitions (2.16-2.21)
of section 2 and took the large $r$ limit (using {\sl Maple} and {\sl GR tensor}). The
result was found to be [6]
$$M_{(5)}=\sqrt{1-{4\tilde\alpha\over
l^2}}\left[{m\pi\over4(\Xi_a\Xi_b)^2}(2\Xi_a+2\Xi_b-\Xi_a\Xi_b)\right]\ , \
J_{(5)a}=\sqrt{1-{4\tilde\alpha\over l^2}}\left[{\pi
ma\over2\Xi_a^2\Xi_b}\right]\,.\eqno(4.1)$$ 
We calculated similarly the mass and
angular momentum in $D=6,7,8$ dimensions in the case when there is only one non
zero rotation parameter. The results could be cast under the form ($D=5,6,7,8$) [6]
 $$M_D=\sqrt{1-{4\tilde\alpha\over l^2}}\,
{m {\cal V}_{D-2}\over4\pi\Xi_a^2}\left[1+{(D-4)\over2}\,\Xi_a\right] \quad,\quad
J_{D}=\sqrt{1-{4\tilde\alpha\over l^2}}\, {ma {\cal
V}_{D-2}\over4\pi\Xi_a^2}\eqno(4.2)$$
where ${\cal V}_{D-2}$ is the volume of the
$(D-2)$-unit sphere. Finally, we extrapolated to all dimensions the asymptotic
expressions for the Kerr-AdS Riemann tensor obtained with {\sl Maple} and {\sl GR
tensor} in $D=5,6,7,8$ dimensions,
and found
(by hand) the mass and angular momenta of the general D-dimensional Kerr-like
solutions of Einstein-Gauss-Bonnet theory as [6]
$$
M_{2n}=\sqrt{1-{4\tilde\alpha\over l^2}}\left[{m{\cal
V}_{2n-2}\over4\pi\Xi}\sum_{i=1}^{i=n-1}{1\over\Xi_i}\right]\quad,\quad
M_{2n+1}=\sqrt{1-{4\tilde\alpha\over l^2}}\left[{m{\cal
V}_{2n-1}\over4\pi\Xi}\left(\sum_{i=1}^{i=n}{1\over\Xi_i}-{1\over2}\right)\right]$$
$$\quad\quad J_{(D)i}=\sqrt{1-{4\tilde\alpha\over l^2}}\left[{m{\cal
V}_{D-2}\over4\pi\Xi}{a_i\over\Xi_i}\right]\,,\eqno(4.3)$$ 
where $\Xi=\Pi_{i=1}^{i=(n-1)}\,\Xi_i$ for $D=2n$ and $\Xi=\Pi_{i=1}^{i=n}\,\Xi_i$
for
$D=2n+1$.

\bigskip
Let us fill up some details of the calculation.

We worked in Kerr-Schild coordinates in which the metric is given for example by
(3.6-3.8). Some useful properties are that the metric coefficients do not depend on
$x^0$ and $\phi_i$; the $g_{\mu_i\mu_j}$ coefficients do not depend on $m$; the other
off-diagonal components are linear in $m$ and hence do not enter the calculation.
The fact that the vector $h^\mu$ is null further simplifies the calculation since
$\sqrt{-g}=\sqrt{-\bar g}$. Using those informations one thus arrives for example at
$$D^{[0}\hat\xi_t^{1]}-\overline{D^{[0}\hat\xi_t^{1]}}=\overline{D^{[0}\hat\xi_t^{1]}}
\left({\delta g\,'_{tt}\over\overline{g}\,'_{tt}}-{\delta g_{tt}\over\overline{g}_{tt}}
-{\delta g_{rr}\over\overline{g}_{rr}}\right)+{\cal O}(m^2)\quad\hbox{with}\qquad
\overline{D^{[0}\hat\xi_t^{1]}}=-{\sqrt{-\overline{g}\,}\ \overline{g}\,'_{tt}\over2
\overline{g}_{tt}\overline{g}_{rr}}\eqno(4.4)$$
where $\xi^\mu_t=(1,0,...,0)$, where
$\delta g_{\mu\nu}\equiv
g_{\mu\nu}-\overline{g}_{\mu\nu}$ and where a prime denotes differentiation with
respect to $r$. A similar expression is obtained for $\hat\xi_t^{[0}k_E^{1]}$ with
$k_E^\mu$ given by equation (2.19) of section 2.

Using then the asymptotic form of the D-dimensional Kerr-AdS metric found in [4] at
leading order in $r$  one then arrives for example at
$$\hat\xi_t^{[0}k_E^{1]}=-{m\over\Xi}\sqrt{g_{(D-2)}}+{\cal
O}\left({m^2\over r}\right)\,,
\eqno(4.5)$$
and
$$
\hat J^{[01]}_{E\, t}={m\over 8\pi\Xi}\,[(D-1)W-1]\sqrt{g_{(D-2)}}+{\cal
O}\left({m^2\over
r}\right)\quad\hbox{with}\quad W\equiv\sum\limits_{i=1}^{i=n}{\mu_i^2\over\Xi_i}
\eqno(4.6)
$$ 
where $g_{(D-2)}$ is the determinant of  the metric $d\sigma^2_{(D-2)}$ in (3.9) with
tildes dropped and with $a_n=0$ in even dimensions $D=2n$.

The calculation of $\hat\xi_t^{[0}k_{GB}^{1]}$ is straightforward since
$$k^\mu_{GB}=-{2\over{\cal L}^2}(D-2)(D-3)\, k^\mu_E+{\cal O}(m^2)\,.\eqno(4.7)$$

 The calculation of $-4\pi
\hat J^{[01]}_{P,t}\equiv(P^{01\alpha\beta}D_{[\alpha}\hat\xi_{t\,\beta]}-
\overline{P^{01\alpha\beta}D_{[\alpha}\hat\xi_{t\,\beta]}})$ at leading order yields in
a first step
$$-4\pi
\hat J^{[01]}_{P,t}=
{\sqrt{g_{(D-2)}}\over\Xi{\cal L}^2}\left[-{2\over{\cal
L}^2}\lim_{r\to\infty}r^{D+1}\delta R^0_{\ 101}
+mW(D-1)(D-2)(D-3)-2m(D^2-5D+8)\right]\eqno(4.8)$$
and requires the knowledge
of the asymptotic form of the Riemann tensor. The structure of the relevant
component for the calculation of the mass
 is easily guessed and depends on two constants only :
$$\delta R^0_{\ 101}\equiv\lim_{r\to\infty}(R^0_{\ 101}-\overline{R^0_{\
101}})={m{\cal L}^2\over r^{D+1}}(D-1)[c_1(D-3)W-c_2]\eqno(4.9)$$
 where, recall,
$W\equiv\sum_{i=1}^{i=[D/2]}{\mu_i^2\over\Xi_i}$, with
$\sum_{i=1}^{i=[D/2]}\mu_i^2=1$ and $\Xi_{D/2}=1$ if $D$ is even. The {\sl Maple} and
{\sl GRtensor} calculations in
$D=5$ determine the constants $c_1$ and $c_2$ to be one, similar calculations in
$D=6,7,8$
 confirm them, so that  the results can confidently be extended to any
$D$. 

After gathering all the previous information the remaining task is to integrate
the superpotential on the sphere at infinity. While the ``multi-cylindrical"  pairs of
coordinates
$(\mu_i,\phi_i)$ were
convenient in [4]  to find the Kerr-anti-de Sitter metric as well as in our previous
calculations, the explicit integrations are more
easily performed
 in spherical coordinates, say, $\varphi_k$ with $(k=1, 2,\cdots,
D-2)$.

\medskip
The final results are given above, formulae (3).

\medskip
The calculation of the angular momenta proceeds along similar lines. Note that neither
the background nor the vectors $k^\mu_E$ and $k^\mu_{GB}$ enter the calculations.
However, in Einstein-Gauss-Bonnet theory, they are not given simply by Komar's
integrals because the first term in (2.18) (that is $\hat J^{[01]}_{P,i}$) is not zero.

\section{Conclusion}

Formulae (4.3) for the mass and angular momenta of Kerr-AdS spacetimes in
Einstein-Gauss-Bonnet theory 
 have the right
limits. When $\alpha=0$ (Einstein's theory) they reduce to the results obtained in [2]
and [5] and,  as shown in [2], they are consistent with the first law of Einstein black
hole thermodynamics.

When there is no rotation ($a_i=0, \Xi_i=1$) the masses (4.3) reduce to the
result obtained in [12-13] and [3] (noting that the coefficients of the metric (3.6)
then tend to $g_{tt}\simeq 1/g_{rr}\simeq r^2/{\cal
L}^2-2m/r^{D-3}$) and are also consistent with the first law of Einstein-Gauss-Bonnet
black hole thermodynamics (where the entropy is no longer proportional to the
surface area of the horizon, see e.g. [12-13]).

Since the overall factor in (4.3) is the
same as the one which appears in (3.4) one may conjecture that in the general
Lovelock theory in $D$ dimensions (whose Lagrangian is the sum of the first
$[D/2]$ dimensionally continued Euler forms) the expressions for the masses and
angular momenta will also be proportional to their values in Einstein's theory
with an overall coefficient which is zero when the equation for the maximally
symmetric space curvature has one single root of maximal multiplicity $[D/2]$.

Finally, the decisive check of formulae (4.3), that is of the pertinence of
 the proposal (2.20) made in [3] for the vector $k^\mu_{GB}$, will be possible when
the full geometry of the Kerr-like rotating {\it black holes} is known, by seeing
if, with such definitions, the first law of thermodynamics is still satisfied. 

\section{Acknowledgements}
Most of the work summarized here is the result of greatly enjoyable collaborations  with Joseph Katz, Yoshiyuki Morisawa
 and Sachiko Ogushi.

\end{document}